\documentclass[conference]{IEEEtran}
\IEEEoverridecommandlockouts
\usepackage[left=1.62cm,right=1.62cm,top=1.9cm]{geometry}
\usepackage{xcolor,soul,framed} 

  \IEEEoverridecommandlockouts
    \usepackage{cite}
    \usepackage{amsmath,amssymb,amsfonts}
    \usepackage{algorithmic}
    \usepackage{graphicx}
    \usepackage{textcomp}
    \usepackage{xcolor}
    \def\BibTeX{{\rm B\kern-.05em{\sc i\kern-.025em b}\kern-.08em
    T\kern-.1667em\lower.7ex\hbox{E}\kern-.125emX}}
    
\usepackage{cite}
\usepackage{amsmath,amssymb,amsfonts}
\usepackage{algorithmic}
\usepackage{graphicx}
\usepackage{textcomp}
\usepackage{xcolor}
\usepackage{etoolbox}
\makeatletter
\patchcmd{\@makecaption}
  {\scshape}
  {}
  {}
 {}
\makeatother

\def\BibTeX{{\rm B\kern-.05em{\sc i\kern-.025em b}\kern-.08em
    T\kern-.1667em\lower.7ex\hbox{E}\kern-.125emX}}
\colorlet{shadecolor}{yellow}
\graphicspath{{../pdf/}{../jpeg/}}
\DeclareGraphicsExtensions{.pdf,.jpeg,.png}

\usepackage{array}
\usepackage{mdwmath}
\usepackage{mdwtab}
\usepackage{eqparbox}
\usepackage{url}
\usepackage{amsmath}
\let\labelindent\relax
\usepackage{enumitem}
\usepackage{tabularx,booktabs}
\newcolumntype{C}{>{\centering\arraybackslash}X} 
\usepackage{lipsum} 

\makeatletter
    \newcommand{\linebreakand}{%
      \end{@IEEEauthorhalign}
      \hfill\mbox{}\par
      \mbox{}\hfill\begin{@IEEEauthorhalign}
    }
    \makeatother
    
\begin{document}

    \title{Navigating AI Policy Landscapes: Insights into Human Rights Considerations Across IEEE Regions\\
    }

\author{
  \IEEEauthorblockN{1st Angel Mary John\thanks{Personal use of this material is permitted.  Permission from IEEE must be obtained for all other uses, in any current or future media, including reprinting/republishing this material for advertising or promotional purposes, creating new collective works, for resale or redistribution to servers or lists, or reuse of any copyrighted component of this work in other works. Published article \cite{john2024navigating}}}
  \IEEEauthorblockA{\textit{Dept. of Law} \\
    \textit{Mar Gregorios College of Law}\\
    Trivandrum, India \\
    angel@mgcl.ac.in}
  \and
  \IEEEauthorblockN{2nd Jerrin Thomas Panachakel}
  \IEEEauthorblockA{\textit{Dept. of Electronics and Comm. Engg.} \\
    \textit{College of Engineering}\\
    Trivandrum, India \\
    jerrin.panachakel@cet.ac.in}
  \and
  \IEEEauthorblockN{3rd Anusha S.P.}
  \IEEEauthorblockA{\textit{Dept. of Civil Engg.} \\
    \textit{College of Engineering}\\
    Trivandrum, India \\
    anusha@cet.ac.in}
\thanks{Financial assistance from APJ Abdul Kalam Technological University, Kerala, India, under the scheme for providing financial assistance to faculty members for presenting research papers at international conferences was received for presenting this work.}
}


\maketitle

\textbf{This paper explores the integration of human rights considerations into AI regulatory frameworks across different IEEE regions—specifically the United States (Region 1-6), Europe (Region 8), China (part of Region 10), and Singapore (part of Region 10). While all acknowledge the transformative potential of AI and the necessity of ethical guidelines, their regulatory approaches significantly differ. Europe exhibits a rigorous framework with stringent protections for individual rights, while the U.S. promotes innovation with less restrictive regulations. China emphasizes state control and societal order in its AI strategies. In contrast, Singapore's advisory framework encourages self-regulation and aligns closely with international norms. This comparative analysis underlines the need for ongoing global dialogue to harmonize AI regulations that safeguard human rights while promoting technological advancement, reflecting the diverse perspectives and priorities of each region.}



\section{Introduction}
Artificial intelligence (AI) is a transformative technology that is reshaping many aspects of society, economy, and governance. By simulating human intelligence processes through computer systems and software, AI has the potential to optimize tasks ranging from the mundane to the complex, across various sectors including healthcare, finance, and education. However, as AI systems become more integrated into daily activities and critical infrastructures, they also pose new challenges and ethical considerations, particularly concerning human rights \cite{john2023ethical}.

The evolution of technology has historically posed challenges and opportunities for human rights \cite{kirby1988human}. With AI, these challenges are magnified due to the scale, speed, and complexity of the technology. Effective AI regulations require a nuanced understanding of both the technical aspects of AI systems and the socio- political contexts in which these systems operate. 

Human rights in the context of AI are crucial because these technologies can significantly affect fundamental rights such as privacy and freedom of expression \cite{brownsword2022ai}. The pervasive data collection capabilities of AI systems can lead to invasive surveillance, while algorithmic decision-making can perpetuate biases and discrimination if not properly overseen. The intersection of AI and human rights centers on the balance between harnessing the benefits of AI technologies and mitigating their risks to human dignity and freedom. For instance, while AI can enhance access to information and boost learning outcomes, it can also be used for surveillance and social scoring, which can infringe on privacy and other civil liberties \cite{elliott2022ai}. Therefore, understanding how AI impacts human rights is essential to developing regulations that not only foster innovation but also protect individuals and uphold democratic values 

This paper aims to provide a comprehensive review of how different jurisdictions—namely the United States, China, and Europe—are incorporating human rights considerations into their AI regulatory frameworks. By analyzing these diverse approaches, the paper seeks to understand the global landscape of AI regulation and to highlight best practices that can be adopted to protect human rights in the digital age.
\section{What are Human Rights?}
Human rights constitute the fundamental rights and freedoms that every individual is entitled to from birth to death \cite{marks2014human}. These rights are rooted in universal values such as dignity, fairness, equality, respect, and independence. Defined by law through international treaties, national legislation, and other sources, human rights help to protect individuals against abuses by the state and other institutions, ensuring people can participate in society without discrimination and oppression \cite{donnelly2013universal}.

Human rights are inalienable and universal, meaning they cannot be taken away and must apply universally to all human beings. The Universal Declaration of Human Rights (UDHR), ratified by the United Nations General Assembly in 1948, represents the initial legal codification of essential human rights to be safeguarded globally \cite{alfredsson2023universal}. 

\section{Legal Frameworks for Human Rights}

\subsection{Constitution of the United States on Human Rights}
The Constitution of the United States, complemented by its amendments, especially the Bill of Rights, enshrines numerous protections that align with human rights principles. The First Amendment secures freedoms related to religion, expression, assembly, and the right to petition, prohibiting Congress from favoring one religion over another or hindering an individual's religious practices. The Fourth Amendment protects citizens from unreasonable searches and seizures, and the Eighth Amendment prohibits cruel and unusual punishment. These rights aim to preserve individual dignity and liberty against overreach by the government \cite{madison2021constitution, taylor2022constitution}.

\subsection{European Union's Charter of Fundamental Rights}
In Europe, the Charter of Fundamental Rights of the European Union specifies a range of human rights covered under EU law \cite{eu2012charter}. Adopted in 2000 and legally binding since 2009 with the Treaty of Lisbon, the Charter compiles all personal, civic, political, economic, and social rights enjoyed by people within the EU \cite{kokott2012charter}. The document is significant as it includes rights that are not found in the UDHR, such as the right to data protection and the guarantee of bioethical principles in medicine and biology \cite{oliveira2004right}. These additions reflect modern issues, showcasing the EU’s progressive approach to human rights.

\subsection{Constitution of the People's Republic of China on Human Rights}
China’s approach to human rights within its constitution includes both classic rights and duties of citizens but places them within the context of a socialist system \cite{hualing2018makes}. The Constitution of the People’s Republic of China guarantees many of the usual rights, such as freedom of speech, press, assembly, association, procession, and demonstration. However, these rights are often circumscribed by the state’s prerogative to limit them in the interest of the national security, societal harmony, or community welfare, reflecting a collectivist approach rather than an individualist perspective \cite{deva2011constitution}. Thus, the enforcement and interpretation of these rights can be significantly different from their counterparts in the West \cite{zhang2012constitution}.

\subsection{Constitution of Singapore on Human Rights}
The Constitution of Singapore establishes a structured framework for the protection of human rights, intertwining these protections with the broader principles of public order and national security \cite{tan2015constitution}. Fundamental liberties, including the right to life, personal liberty, and freedom from slavery and forced labor, are firmly enshrined in Part IV of the Constitution. Moreover, it grants citizens the rights to freedom of speech, assembly, and association, as articulated in Article 14. However, these rights are not absolute and carry provisions that allow for restrictions in the interest of security, public order, or morality \cite{thio2009protecting}. 

\section{Evolution of AI}
In the annals of Artificial Intelligence (AI) development, certain foundational systems have been instrumental in shaping the trajectory of the field. Among these, the Logic Theorist, conceived by Allen Newell and Herbert Simon in 1956, represents a seminal advancement. This computational program was engineered to automate the proof of mathematical theorems through the application of symbolic logic, thereby establishing a significant milestone in the evolution of AI \cite{newell1956logic}. Subsequently, the introduction of the General Problem Solver (GPS) by Newell et al. in 1957 marked a further evolution in AI capabilities. This system demonstrated the potential for a computational approach to address a broad array of problems via a strategic exploration of potential solutions. The perceptron, developed by McCulloch and Pitts in 1943, constitutes another cornerstone in the development of AI, introducing the concept of a neural network capable of pattern recognition, an early precursor to modern machine learning techniques \cite{rosenblatt1957perceptron}. Moreover, the creation of ELIZA by Joseph Weizenbaum in 1966 showcased the feasibility of simulating conversational exchanges between humans and computers through natural language processing, further broadening the scope of AI applications \cite{weizenbaum1966eliza}. The ensuing years have witnessed an exponential growth in the AI field, characterized by the continuous emergence of innovative systems, each contributing to the expansion and depth of AI research and application.

Some of the state-of-the-art AI systems are:
\begin{itemize}
    \item \textbf{Generative AI in NLP (Natural Language Processing)}: Generative AI in NLP is capable of producing text that closely mimics human writing and is utilized across various applications from writing assistance to more complex coding tasks \cite{jo2023promise, orru2023human, katz2024gpt}.
    \item \textbf{Autonomous Navigation}: Autonomous driving systems use AI to interpret sensory information to detect surroundings and navigate safely with little or no human intervention. This includes real-time decision-making in traffic, obstacle recognition, and route optimization \cite{muhammad2020deep,soori2023artificial}.
    \item \textbf{Brain-Computer Interfaces}: AI algorithms are used to interpret patterns in neural recordings to decode the words that a person is imagining or the movement the person is intending. This technology holds significant promise for helping individuals with disabilities \cite{panachakel2020novel,panachakel2020improved}.
    \item \textbf{Biomedical Imaging}: AI models are extensively used to analyze medical images such as X-rays, MRIs, and CT scans to detect anomalies like tumors, fractures, or diseases at a much faster rate and with potentially higher accuracy than human radiologists \cite{mamoshina2016applications, suzuki2017overview}.
    \item \textbf{Drug Discovery}: AI is used to predict how different patients will respond to drugs based on their genetic information. This application is crucial for personalizing medicine, potentially reducing side effects and improving drug efficacy \cite{blanco2023role,mak2023artificial}.
    \item \textbf{Digital Assistants}: AI-based Digital Assistants supports a variety of tasks from setting reminders and playing music to providing real-time information like weather forecasts and traffic updates. The technology behind these include a complex mix of voice recognition, understanding natural language, and contextual awareness to provide a personalized user experience \cite{maedche2019ai}.
\end{itemize}
\section{Current AI Regulations}
\subsection{Executive Order on AI in the USA}
In October 2023, President Biden signed a pivotal executive order concerning the Safe, Secure, and Trustworthy Development and Use of Artificial Intelligence. This directive is designed to guide the evolution and implementation of artificial intelligence (AI) in the United States, focusing heavily on human rights. It represents a crucial move towards tackling the intricate issues associated with AI technologies, especially concerning privacy, bias, transparency, and accountability.

Key elements of the executive order:
\begin{itemize}
    \item \textbf{Promotion of AI Research and Development}: The executive order encourages the advancement of AI technologies that are aligned with American values, including respect for civil liberties and human rights. It advocates for federal investments in AI research and development that prioritize ethical considerations.
    \item \textbf{Enhancing AI Governance}: To ensure that AI development is safe, secure, and trustworthy, the order directs federal agencies to adopt governance frameworks that incorporate human rights principles. This includes guidelines for AI deployment in sensitive areas such as surveillance, law enforcement, and decision-making processes that impact civil rights.
    \item \textbf{Bias and Discrimination Mitigation}: A major focus of the order is to mitigate bias and discrimination in AI systems. It mandates the creation of standards and testing procedures to identify and eliminate biases that could lead to unequal treatment under the law, particularly in critical domains like healthcare, employment, and criminal justice.
    \item \textbf{Public Engagement and Transparency}: The order emphasizes the importance of transparency and public engagement in AI policymaking. It calls for open dialogues with stakeholders, including the tech industry, academia, civil society, and the general public, to ensure that AI policies reflect a broad range of perspectives and are communicated clearly to all Americans.
\end{itemize}
\subsection{The Proposed AI Act of the European Union}

The European Union's proposed AI Act represents a significant stride in creating a comprehensive legal framework to govern the use and development of artificial intelligence technologies. It is designed to address the various risks associated with AI while promoting innovation and upholding the EU's digital and ethical standards.

Key elements of the proposed act:
\begin{itemize}
    \item \textbf{Risk-Based Classification System}: The AI Act introduces a risk-based approach to AI regulation, categorizing AI systems into four levels of risk: unacceptable risk, high risk, limited risk, and minimal or no risk. This classification dictates the regulatory requirements for each AI system, focusing particularly on high-risk and unacceptable-risk applications.
    \item \textbf{Prohibitions of Certain AI Practices}: The AI Act identifies specific uses of AI that are considered harmful and prohibits them. This includes practices like social scoring by governments that could lead to discrimination or the use of AI for manipulative or exploitative purposes.
    \item \textbf{Requirements for High-Risk AI Applications}: High-risk AI applications, such as those involved in critical infrastructures, education, employment, and law enforcement, are subject to stringent obligations before they can be brought to market. These requirements include conducting adequate risk assessments, ensuring data governance, maintaining detailed documentation, and implementing robust human oversight.
    \item \textbf{Transparency Obligations}: The AI Act mandates specific transparency requirements for certain AI systems, particularly those that interact directly with users, such as chatbots. Users must be informed that they are interacting with an AI to ensure they can make knowledgeable decisions about their engagement.
    \item \textbf{Conformity Assessments and Enforcement}: AI providers must conduct conformity assessments to demonstrate compliance with the Act before their AI systems can be marketed or put into service. The Act also outlines enforcement mechanisms and penalties for non-compliance, ensuring that AI systems adhere to the regulations continuously.
    \item \textbf{Governance and Oversight}: The AI Act establishes a governance structure at both the European and national levels to oversee the implementation of the regulations. This structure is responsible for ensuring compliance, managing risks, and maintaining a register of high-risk systems.
\end{itemize}
\subsection{AI Regulation in China}
China has actively been developing a comprehensive framework for the regulation of artificial intelligence, reflecting its ambitions to be a global leader in AI technologies. The Chinese government views AI as a pivotal element in its broader strategy for technological and economic development. The regulatory framework for AI in China is designed to address both the promotion of AI technologies and the management of associated risks.

Key Aspects of AI Regulation in China:
\begin{itemize}
    \item \textbf{National Strategies and Policies}: China has issued several key policy documents that outline the country's ambitions in AI. The "New Generation Artificial Intelligence Development Plan" (AIDP), released in 2017, sets broad goals for achieving global leadership in AI by 2030. This plan emphasizes the integration of AI in industry, military, and societal contexts, aiming to create a massive AI industry and improve public security and governance through AI technologies.
    \item \textbf{Ethical and Safety Standards}: Recognizing the potential risks associated with AI, China has begun to implement standards and guidelines to ensure the ethical use of AI. These include specifications on the ethical norms for AI, the management of data in AI applications, and safety requirements. For example, in 2019, China's Standardization Administration released guidelines that include principles such as fairness, transparency, and respect for user privacy.
    \item \textbf{Sector-Specific Regulations}: China has been developing sector-specific regulations for AI, particularly in areas like finance, healthcare, and transportation. These regulations aim to ensure that AI applications in these fields are safe, reliable, and compliant with existing laws and standards.
    \item \textbf{Security and Control}: The Chinese approach to AI regulation also strongly emphasizes national security and social stability. Regulations are likely to include mechanisms for government oversight and control of AI technologies, particularly those that might impact public opinion or social order, such as algorithms used in social media and public surveillance systems.
\end{itemize}
\subsection{Model AI Governance Framework of Singapore}

Singapore's regulatory approach to artificial intelligence is encapsulated in its Model AI Governance Framework, which is part of the country's broader strategy to promote the ethical use of AI technologies while fostering innovation. The framework highlights Singapore's proactive stance on AI governance and reflects its commitment to establishing a trustworthy AI ecosystem.

Key Aspects of AI Regulation in Singapore:

\begin{itemize}
    \item \textbf{Comprehensive Guidelines}: Singapore's framework provides detailed guidance for the responsible deployment of AI technologies. It is designed to be practical, user-friendly, and adaptable across different industries, encouraging organizations to implement AI responsibly while considering ethical implications.
    \item \textbf{Ethical Principles}: The framework emphasizes AI systems being explainable, transparent, and fair. This approach ensures that AI-assisted decision-making upholds values of accountability and fairness, which are crucial for maintaining public trust.
    \item \textbf{Human-Centric Approach}: The guidelines advocate for AI systems to enhance human capabilities and safeguard human interests, ensuring that the deployment of AI technologies does not compromise individual rights and welfare.
    \item \textbf{Implementation and Self-Assessment Guide for Organisations (ISAGO)}: This guide helps organizations assess how their AI governance practices align with the Model Framework. It provides a self-assessment checklist and best practices, which help organizations measure and enhance their compliance with the framework.
    \item \textbf{Compendium of Use Cases}: The Compendium illustrates how various local and international companies have successfully implemented the Model Framework. This resource serves as a practical reference for organizations looking to adopt similar AI governance practices.
\end{itemize}

\section{Comparison of AI Regulations: US, Europe, China, and Singapore}

The regulatory landscapes for artificial intelligence (AI) across the US, Europe, China, and Singapore showcase both convergences and divergences in approach, reflecting different priorities, cultural contexts, and levels of regulatory maturity.

\subsection{Similarities in AI Regulation}
Across these regions, several common themes emerge in AI regulation:
\begin{itemize}
    \item \textbf{Ethical Focus}: All four regions emphasize the ethical deployment of AI technologies, focusing on principles such as fairness, accountability, and transparency. This is evident in Europe's AI Act, Singapore's Model AI Governance Framework, the ethical guidelines published by the US, and China's developing standards for ethical AI.
    \item \textbf{Risk-Based Regulatory Approach}: The US, Europe, China, and Singapore have adopted or are moving towards a risk-based approach to AI regulation, categorizing AI applications based on their potential risk to society and tailoring regulatory requirements accordingly. This approach aims to mitigate risks without stifling innovation.
    \item \textbf{Public Engagement and Stakeholder Involvement}: There is a shared emphasis on engaging various stakeholders in the regulatory process. Public consultations, multi-stakeholder committees, and partnerships with academia and industry are common strategies to ensure that AI regulations are well-informed and effective.
    \item \textbf{International Collaboration}: Despite regional differences, there is a recognition of the need for international cooperation in AI regulation. Europe, the US, China, and Singapore have all expressed interest in participating in global discussions to harmonize AI standards and practices.
\end{itemize}

\subsection{Differences in AI Regulation}
Despite these similarities, significant differences in AI regulatory frameworks reflect divergent approaches and priorities:
\subsubsection{Scope and Stringency of Regulations}

The scope and stringency of AI regulations vary significantly across the US, Europe, China, and Singapore, reflecting their diverse approaches to managing the balance between innovation and control. Europe's AI Act is notably comprehensive and stringent, setting detailed requirements for high-risk AI applications and explicitly banning certain AI practices deemed too risky. This approach is aligned with Europe's strong emphasis on individual rights and privacy, as seen in other regulations like the GDPR.

In contrast, the US adopts a more sector-specific and fragmented approach. AI regulation in the US tends to focus on promoting innovation and maintaining competitiveness in the global technology market. Regulatory measures are often voluntary, relying on guidelines rather than strict legislative mandates, which allows for greater flexibility but might lead to inconsistent applications across different sectors and states.

China's regulatory approach integrates AI governance with broader national strategies for technology and security. The Chinese government exercises strong control over AI development, with a focus on ensuring that AI technologies support state security and social management objectives. This leads to a regulatory environment that can be both broad in its ambitions and strict in its control, particularly concerning surveillance and data handling.

Singapore offers a unique model that blends regulatory oversight with industry facilitation. The Model AI Governance Framework in Singapore is not legally binding but provides comprehensive guidelines that encourage organizations to adopt responsible AI practices voluntarily. This approach reflects Singapore's role as an international business hub and its desire to be a thought leader in AI governance, promoting a balance between innovation and ethical considerations.
  
\subsubsection{Enforcement Mechanisms}
The enforcement mechanisms for AI regulations also differ significantly across regions, reflecting the varying legal and governance structures of each. In Europe, the AI Act includes strong enforcement provisions, with the potential for substantial fines similar to those under the GDPR for non-compliance. This reflects Europe's systematic approach to regulation across digital and data protection domains, ensuring that the rules have real teeth to enforce compliance.

In contrast, the United States has a more decentralized approach to enforcement, primarily relying on existing regulatory frameworks and sector-specific guidelines. Enforcement may vary significantly from one state to another and from one industry to another, lacking a unified federal standard. This could potentially lead to inconsistencies in how AI regulations are applied and enforced across the country.

Singapore's enforcement mechanisms are built around its Model AI Governance Framework, which is advisory rather than compulsory. The emphasis is on self-regulation with a focus on capacity building within organizations to understand and implement the guidelines effectively. While this approach promotes flexibility and innovation, it relies heavily on industry cooperation and ethical self-commitment to governance standards.

China’s approach to enforcement involves tight government control and oversight, with specific regulations that align with broader social management and national security goals. Enforcement mechanisms are likely stringent, reflecting the central government’s role in regulating and controlling the technology sector, particularly in areas considered critical to national interests.

\subsubsection{Cultural and Political Contexts}
The cultural and political contexts significantly influence the development and implementation of AI regulations in each region. In China, the government's approach to AI regulation is intertwined with its socio-political agenda, emphasizing surveillance, social stability, and control. The regulations are designed to bolster government oversight and integrate AI technologies into a broader socio-political framework, which supports state surveillance and social governance.

In Europe, the emphasis is on protecting individual rights and privacy, reflecting a cultural and political context that values democratic governance and the protection of personal freedoms. The AI Act is part of a broader movement towards ensuring technology serves the public good without compromising personal integrity or democratic values.

The US approach to AI regulation is shaped by its liberal market economy, prioritizing innovation and the protection of intellectual property within a competitive global market. The regulatory framework is more about fostering innovation and less about imposing controls, reflecting a cultural ethos that values entrepreneurship and technological leadership.

Singapore's model reflects its status as a global business hub, aiming to establish standards that foster trust and safety in AI applications while also promoting the city-state as a center for technology and innovation. The regulatory approach is pragmatic, focusing on maintaining a competitive edge in the global market while ensuring that AI development aligns with public interest and ethical standards.

\section{Conclusion}

This paper has undertaken a comprehensive examination of the integration of human rights considerations into the regulatory frameworks for artificial intelligence (AI) across different jurisdictions, with a particular focus on the United States, China, Europe, and Singapore. Each region presents a unique blend of strategies shaped by its socio-political context, priorities, and cultural values, which influence how AI technologies are governed to align with human rights norms.

The comparison reveals that while there is a universal acknowledgment of the potential of AI to both enhance and challenge human rights, the approaches to harnessing its benefits while mitigating its risks vary significantly. Europe stands out with its detailed and stringent regulatory framework, emphasizing the protection of fundamental rights and transparency. The United States, by contrast, adopts a more decentralized and innovation-friendly approach that prioritizes sector-specific guidelines and fosters technological leadership. Singapore’s framework, advisory in nature, emphasizes ethical deployment through industry collaboration and self-regulation, reflecting its status as a global business hub seeking to lead in AI governance. China's strategy is tightly controlled, integrating AI regulation within broader national security and social management goals, thus ensuring state oversight over AI development and deployment.

This global landscape of AI regulation underscores the need for an ongoing dialogue between nations to share best practices and strive for some level of harmonization in regulations, especially as AI technologies continue to evolve at a rapid pace. Such international collaboration could foster global standards that not only propel innovation but also safeguard human dignity, privacy, and rights in the digital age.

Ultimately, as AI technologies become increasingly embedded in every aspect of human life, the continuous evolution of regulatory frameworks will be crucial. They must be adaptable to emerging AI advancements and sensitive to the dynamic interplay between technology and human rights. This balance is essential not only for protecting individuals but also for ensuring that AI serves the broader interests of society globally.
\section*{Acknowledgment}
 The authors  extend their sincere gratitude to APJ Abdul Kalam Technological University, Kerala, India, for their generous financial support, including travel assistance, under the scheme for providing financial assistance to faculty members for presenting research papers at international conferences.
\bibliographystyle{IEEEtran}
\bibliography{IEEEabrv,Bibliography}

\vfill

\end{document}